\documentclass[aps,pra,preprint,superscriptaddress,amsmath,amssymb,showpacs]{revtex4-2}
\usepackage[english]{babel}
\usepackage{amssymb,bm}
\usepackage{graphicx}
\usepackage{amsmath}
\usepackage{color}
\usepackage{comment}
\usepackage[colorlinks=true,citecolor=blue,urlcolor=blue]{hyperref}
\begin{document}
 \begin{titlepage}

\title{Electron scattering by a magnetic monopole in solid-state experiments}
\author{P. S. Sidorov}\email{pavel.sidorov@metalab.ifmo.ru}
\affiliation{School of Physics and Engineering, ITMO University, 197101 St. Petersburg, Russia}
\author{N. A. Vlasov}\email{nikolai.vlasov@metalab.ifmo.ru}
\affiliation{School of Physics and Engineering, ITMO University, 197101 St. Petersburg, Russia}
\author{I. S. Terekhov}\email{i.s.terekhov@gmail.com}
\affiliation{School of Physics and Engineering, ITMO University, 197101 St. Petersburg, Russia}
\affiliation{Budker Institute of Nuclear Physics of SB RAS, 630090 Novosibirsk, Russia}
\author{A. I. Milstein}\email{a.i.milstein@inp.nsk.su}
\affiliation{Budker Institute of Nuclear Physics of SB RAS, 630090 Novosibirsk, Russia}\affiliation{Novosibirsk State University, 630090 Novosibirsk, Russia}

\date{\today}

\begin{abstract}
The scheme of experiment for studying electron scattering in the field of a magnetic monopole in two dimensional electron gas is proposed. The differential scattering cross section is obtained in the eikonal approximation. For unpolarized initial electron, the differential  cross section coincides with that in the field of an infinitely long solenoid, up to redefinition  of a magnetic flux. It is shown that  the  scattered electron becomes polarized  even for unpolarized  initial electron. Besides,  in an experimental setup similar to the Hall experiment, the spin polarization arises in the direction perpendicular to the electron current.

\end{abstract}

\maketitle
 \end{titlepage}
\section{Introduction}
The quantum problem of a charged particle  in the field of a magnetic monopole was considered for the first time in Ref.~\cite{Dirac1931}. It was shown that the existence of a magnetic monopole leads to the quantization of electric charge. Although elementary particles with nonzero magnetic charge have not yet been found \cite{Mitsou2025}, structures similar to Dirac magnetic monopoles have been predicted and observed in solid-state experiments, namely in spin ice \cite {Castelnovo2008,Khomskii2014,Chen2016,Alexanian2024}. Materials in which spin ice occurs are dielectrics, so it is difficult to use them to study the electron scattering in the field of a magnetic monopole.  

It is well known that the field of magnetic monopole can be generated using a Dirac string, which is  semi-infinite narrow solenoid.  Therefore, electron scattering in a monopole field can be studied using the following setup. Let us place a solenoid of length $L$ and radius $a$ along the z-axis. One end is placed at $z = 0$ and the other one at $z = -L$. In this case, at $z = 0$ and  $a\ll\rho\ll L$, where $\rho=\sqrt{x^2+y^2}$, the vector potential and magnetic field read  
\begin{align}
	& \bm A(\bm \rho)= \dfrac{\Phi}{4\pi}\frac{[\bm\nu\times\bm \rho]}{\rho^2}, \quad\bm B(\bm \rho)=  \dfrac{\Phi}{4\pi}\frac{\bm \rho}{\rho^3}\,,\label{AandB}
\end{align}
where $\Phi$ is the magnetic flux through the solenoid, $\bm \rho=(x,y)$, and $\bm\nu$ is the unit vector directed along the $z$-axis. The vector potential $\bm A(\bm \rho)$ and  the magnetic field $\bm B(\bm \rho)$ correspond to a magnetic monopole having the magnetic charge  $\Phi/(4\pi)$. By placing a two-dimensional electron gas in the plane  $z=0$, one can study the scattering of electrons in the field of a magnetic monopole.  Thus, although magnetic monopoles have not yet been detected, it is possible to study the scattering of electrons in its magnetic field in solid-state experiments. Furthermore, the magnitude of the magnetic charge can be changed by changing the magnetic flux.

In Ref.~\cite{MT2025_2}, the influence of the finite length of a solenoid on the Aharonov-Bohm effect \cite{AB1959} is investigated. In that work,  the scattering of an electron in the field of two oppositely charged magnetic monopoles separated by a distance $L$ is considered. Using the eikonal approximation, the asymmetry in the differential scattering cross section is obtained.

In the present paper, using the results of \cite{MT2025_2,MT2025}, we investigate the scattering of an electron moving in the $z=0$ plane in the field \eqref{AandB}. Note that the vector potential $\bm A(\bm\rho)$ coincides with the vector potential of an infinitely long and narrow solenoid, up to the factor of $1/2$. It is known that the differential scattering cross section in the Aharonov–Bohm effect is singular at small scattering angles and the total cross section is infinite \cite{AB1959}. This is due to the slow decrease of the vector potential at large distances, $\bm A (\bm \rho)\sim 1/\rho$. In the case of a magnetic monopole, the vector potential decreases also as $1/\rho$. This means that the differential cross section in the monopole field is also singular at small angles, and the total cross section is infinite.  Therefore, in our paper we consider electron scattering at small angles. Note that the contribution to the scattering cross section at large angles comes from the region $\rho\sim a$. In this region, the vector potential and the magnetic field differ significantly from \eqref{AandB}. Furthermore, the scattering cross section for large angles depends on the experimental setup, i.e., on the boundary conditions for the electron wave function at short distances. However, the differential cross section at large angles   is negligible.

To investigate the electron scattering at small angles, we solve the Schr\"odinger equation using the eikonal approximation. In the leading eikonal approximation, the differential scattering cross sections on a magnetic monopole and on an infinitely long solenoid are identical, except for a redefinition of the magnetic flux. However, the account for the next-to-leading eikonal  approximation results in the appearance of a nonzero spin polarization of scattered electrons. This effect is absent in scattering on an infinitely long narrow solenoid.

\section{Eikonal approximation}
The Schr\"odinger equation for an electron in the field of a magnetic monopole is:
\begin{eqnarray}\label{InEq}
	\frac{(\bm p-\frac{e}{c}\bm A)^2}{2m}\psi-\frac{e\hbar\alpha}{2mc} (\bm \sigma \cdot \bm B)\psi=E\psi\,,
\end{eqnarray}
where $\bm p=-i\hbar(\partial/\partial x,\partial/\partial y)$, $\hbar$ is the Plank constant, $\bm\sigma$ are Pauli matrices, $c$ is speed of light, $m$ is the electron mass, the coefficient $\alpha$ takes into account a possible influence of the material on the electron gyromagnetic ratio,  $\bm A$ and $\bm B$ are given in \eqref{AandB}. In Eq.~\eqref{InEq},   the terms with the Dirac delta function $\delta(\bm\rho)$ are omitted, since  the scattering amplitude at small angle are considered. Below we set $\hbar=c=1$.

Let the momentum of an initial electron is directed along the x-axis. To find the wave function in the eikonal approximation \cite{LLQM}, we substitute it in the form $\psi=e^{i k x}F(\rho)$ into Eq.~\eqref{InEq} and obtain
\begin{eqnarray}\label{LeadAppr}
\left(\frac{\partial}{\partial x}+i\gamma\frac{y}{\rho^2}\right)F=\frac{i}{2k}\left[\frac{\partial^2 F}{\partial x^2}+\frac{\partial^2 F}{\partial y^2}-
\frac{2i\gamma}{\rho^2}\left(x\frac{\partial}{\partial y}-y\frac{\partial }{\partial x}\right)F-\frac{\gamma^2}{\rho^2}F+\frac{\alpha\gamma}{\rho^3}(\bm \sigma\cdot\bm \rho)F\right],
\end{eqnarray}
where $k=\sqrt{2mE}$, $\gamma=e\Phi/(4\pi)$. The right-hand side of this equation is suppressed by the factor $1/k$. To find the leading eikonal approximation $F_{1}(x,y)$, we set the right-hand side of Eq.~\eqref{LeadAppr}  to be zero, 
\begin{eqnarray}
	\left(i\frac{\partial}{\partial x}-\gamma\frac{y}{\rho^2}\right)F_1=0.
\end{eqnarray}
The solution of this equation has the form
\begin{eqnarray}\label{LedingSolutionF}
	F_1(x,y)=U_i\exp\left\{-i\gamma\left(\arctan\left(\frac{x}{y}\right)+ \frac{\pi}{2}\text{sign}(y)\right)\right\}\,,
\end{eqnarray}
where $U_i$ is the  spinor of an initial electron. The solution \eqref{LedingSolutionF} coincides with the eikonal that in the Aharonov-Bohm problem \cite{LLQM,MT2025_2}. 
To find the solution in the next-to-leading approximation,  we substitute $F=F_1+F_2$ and hold the terms $\propto F_1$ in the right-hand side of Eq.~\eqref{LeadAppr}, 
\begin{eqnarray}\label{NextToLeadAppr}
	\left(\frac{\partial}{\partial x}+i\gamma\frac{y}{\rho^2}\right)F_2=i\frac{\alpha\gamma }{2k\rho^3}(\bm \sigma\cdot\bm \rho)F_1\,.
\end{eqnarray}
The solution of Eq.~\eqref{NextToLeadAppr} is
\begin{eqnarray}
	F_2(x,y)=-i\frac{\alpha\gamma }{2k\rho}\left(\sigma_x-\frac{x+\rho}{y}\sigma_y\right)F_1(x,y)\,.
\end{eqnarray}
It is seen  that  in the next-to-leading approximation a spinor in the final state differs from that in the initial state.

\section{Scattering problem}
To find the scattering amplitude, we extract the diverging wave, perform standard calculations, see Ref. \cite{LLQM} , and obtain  the wave function at large distances,
\begin{eqnarray}
	\psi(\bm \rho)=e^{i k x}U_i+\frac{e^{ik\rho}e^{-i\pi/4}}{\sqrt{\rho}}\sqrt{\frac{k}{2\pi}}\int d\bm\rho'e^{-i\bm q\bm \rho'}\left[\frac{\partial(F_1+F_2)}{\partial x'}-\frac{i}{2k}\left(\frac{\partial^2 F_1}{\partial x'^2}+\frac{\partial^2 F_1}{\partial y'^2}\right)\right]\,,
\end{eqnarray}
where $\bm q=k\frac{\bm \rho}{\rho}-k\bm e_x\approx k\phi\,\bm e_y$, $\phi$ --
 scattering angle, $\bm e_{x}$ and $\bm e_{y}$ are the unit vectors directed along the
$x$ and $y$ axis, respectively. Taking the integral with the required accuracy in $1/k$ and keeping the  singular terms in the parameter $\phi$, we obtain
\begin{eqnarray}
	\psi(\bm \rho)=e^{i k x}U_i+\frac{e^{ik\rho}}{\sqrt{-i\rho}}\frac{i\sin\pi\gamma}{\sqrt{2\pi k}}\left(\frac{2}{\phi}+\left(\alpha\gamma\ln\phi^2\right)\sigma_y\right)U_i\,.
\end{eqnarray}
Therefore the scattering amplitude reads
\begin{eqnarray}\label{ScattAmpl}
	\hat{F}\approx F_0+\bm F\bm \sigma,
\end{eqnarray}
where
\begin{eqnarray}
	F_0&=&i\frac{2\sin\pi\gamma}{\phi\sqrt{2\pi k}}\,,\label{F0}\\
	\bm F&=&i\frac{\alpha\gamma\sin\pi\gamma\ln\phi^2}{\sqrt{2\pi k}}\bm e_y\,,\label{F1}
\end{eqnarray}
The contribution $F_0$ coincides with the scattering amplitude on an infinitely long solenoid with a magnetic flux  to be $\Phi/2$. The contribution $\bm F$ is absent in the Aharonov-Bohm effect, since the magnetic field outside the infinitely long solenoid is zero. It is clear that $F_0$ is singular at small angles as $1/\phi$. The contribution $\bm F$ appears due to the nonzero magnetic field of the magnetic monopole. It is also singular at small angles, but only as $\ln |\phi|$.

Let us discuss the polarization of scattered electrons. The spinor $U_f$ describing  the final state is 
\begin{align}\label{dec1}
	&U_f=(F_0+ \bm\sigma\cdot\bm F)U_i\,.
\end{align}
The polarization vector $\bm\zeta_f$ of scattered electrons is
\begin{align}
	&	\bm\zeta_f=\dfrac{U_f^\dagger \bm\sigma U_f}{U_f^\dagger U_f}\,.
\end{align}
Performing elementary calculations, we obtain
\begin{align}
	&U_f^\dagger \bm\sigma U_f=[|F_0|^2-|\bm F|^2]\bm\zeta_i+2\mbox{Re}\,(F_0^*\bm F)
	+2\mbox{Re}\,[\bm F^*\,(\bm\zeta_i\cdot\bm F)]+i[\bm F\times\bm F^*]\,,\nonumber\\
	&U_f^\dagger U_f=[|F_0|^2+|\bm F|^2]+2\mbox{Re}\,(F_0^*\bm F)\cdot\bm\zeta_i
	+i[\bm F\times\bm F^*]\cdot\bm\zeta_i\,,
\end{align}
Since $\bm F$ is suppressed relative to $F_0$ by the factor $\phi$ at small $\phi$, we omit the terms quadratic in $\bm F$ and obtain
\begin{align}
	&U_f^\dagger \bm\sigma U_f=|F_0|^2\bm\zeta_i+2\mbox{Re}\,(F_0^*\bm F)
	\,,\nonumber\\
	&U_f^\dagger U_f=|F_0|^2+2\mbox{Re}\,(F_0^*\bm F)\cdot\bm\zeta_i\,.
\end{align}
Thus, the polarization of scattered electrons is nonzero even for unpolarized  initial electrons, i.e. when $\bm\zeta_i=0$,
\begin{align}
	&	\bm\zeta_f=\dfrac{2\mbox{Re}\,(F_0^*\bm F)}{|F_0|^2}\,.\label{Pol}
\end{align}
Substituting of $F_0$ and $\bm F$, see Eqs.~\eqref{F0} and \eqref{F1}, in Eq.~\eqref{Pol}, we obtain
\begin{eqnarray}
	\bm\zeta_f&=&\alpha\gamma\, \phi \ln\left(\phi^2\right)\bm e_y\,. \label{PolResult}       
\end{eqnarray}
Note that the direction of  $\bm\zeta_f$ depends on the sign of $\phi$. Thus, a spin separation occurs in the sample, similar to an experiment studying the spin Hall effect \cite{Dyakov1971}.

In the case of nonzero $\bm\zeta_i$, an asymmetry arises in the differential scattering cross section,
\begin{align}
	&\frac{d\sigma}{d\phi}=|F_0|^2\left(1+\bm\zeta_i\cdot\bm\zeta_f\right)\,.\label{crossInitial}
\end{align}
Substituting $F_0$ and $\bm F$, see Eqs.~\eqref{F0} and \eqref{F1}, in Eqs.~\eqref{Pol} and \eqref{crossInitial}, we obtain
\begin{eqnarray}
	\frac{d\sigma}{d\phi}&=&\frac{2}{\pi k\phi^2}\sin^2(\pi \gamma)\left(1+\bm\zeta_f\cdot\bm\zeta_i\right) \,. \label{crossFinal}       
\end{eqnarray}
For $\bm\zeta_i=0$, the differential cross section coincides with that for scattering on an infinitely long narrow solenoid with the magnetic flux equal to $\Phi/2$.

\section{Conclusion}
The scheme for studying electron scattering in the field of a magnetic monopole is proposed. The "magnetic charge" can be varied by changing the magnetic flux through the solenoid. The differential  scattering cross section is derived in the eikonal approximation. In the case of unpolarized initial electrons, the differential  cross section for small scattering angles coincides with that in the field of an infinitely long narrow solenoid with a magnetic flux equal to $\Phi/2$. Since the monopole magnetic field is nonzero in the region accessible to electron motion, a nonzero polarization $\bm\zeta_f$ of  scattered electrons arises even for $\bm\zeta_i=0$.  In this case, the polarization vector $\bm\zeta_f$ changes sign when replacing $\phi\to-\phi$. This means that there is a spin separation in a sample, similar to an experiment studying the spin Hall effect.

\section*{Acknowledgement}
The work of I.S.T was financially supported by the ITMO Fellowship Program. The work of N.A.V. was supported by the Russian Science Foundation, Grant No. 25-12-00213.


\begin{thebibliography}{99}
	\bibitem{Dirac1931} P. A. M. Dirac, Quantised singularities in the Electromagnrtic Field, Proc. Roy. Soc. Lond. A \textbf{133}, 60 (1931).
	
	\bibitem{Mitsou2025} V. A. Mitsou, J. Phys. Conf. Ser. \textbf{3017}, 012002 (2025). 
	
	\bibitem{Castelnovo2008} C. Castelnovo, R. Moessner, and S. L. Sondhi, Magnetic monopoles in spin ice, Nature  \textbf{451}, 42 (2008).
	
	\bibitem{Chen2016} G. Chen, “Magnetic monopole” condensation of the pyrochlore ice $U(1)$ quantum spin liquid: Application to $\text{Pr}_2\text{Ir}_2\text{O}_7$ and $\text{Yb}_2\text{Ti}_2\text{O}_7$, Phys. Rev. B \textbf{94}, 205107 (2016).
	
	\bibitem{Alexanian2024}Y. Alexanian, J. Saugnier, C. Decorse,  et al. Exploring possible magnetic monopoles-induced magneto-electricity in spin ices,  npj Quant. Mat.. \textbf{9}, 72 (2024).
	
	\bibitem{Khomskii2014} D. Khomskii, Magnetic monopoles and unusual dynamics of magnetoelectrics,  Nat. Commun. \textbf{5}, 4793 (2014).
	
	\bibitem{MT2025_2} A. I. Milstein, I. S. Terekhov, Ann.  Phys.  \textbf{483}, 170263 (2025). 
	
	\bibitem{AB1959} Y. Aharonov and D. Bohm, Phys. Rev. \textbf{115}, 485 (1959).
		
	\bibitem{MT2025} A. I. Milstein, I. S. Terekhov, Phys. Rev. B  \textbf{112}, 024113 (2025).

		
\bibitem{LLQM} L. D. Landau, E. M. Lifshitz, {\it Quantum Mechanics, non-relativistic theory} (Pergamon Press, Oxford, 1977).

\bibitem{Dyakov1971} M. I. Dyakonov and V. I. Perel, Possibility of orientating electron spins with current,  Sov. Phys. JETP Lett. \textbf{13}, 467 (1971).
	

\end{thebibliography}
\end{document}